\documentclass[aps,preprint,supscriptaddress]{revtex4-1}
               
\usepackage{graphicx}   
\usepackage{dcolumn}    
\usepackage{bm}         
\usepackage{amsmath}
\usepackage{amssymb}
\usepackage[usenames]{color}      

\begin{document}

\title{Multiple equilibria in fingering double diffusive convection turbulence}

\author{Yantao Yang}
\affiliation{SKLTCS and Department of Mechanics and Engineering Science, BIC-ESAT,
       College of Engineering, and Institute of Ocean Research, 
       Peking University, Beijing 100871, China}

\author{Roberto Verzicco}
\affiliation{Physics of Fluids Group, Department of Science and Technology, MESA+ Institute, Max Planck Center Twente for Complex Fluid Dynamics, and J. M. Burgers Center for Fluid Dynamics, University of Twente, 7500 AE Enschede, The Netherlands}
\affiliation{Dipartimento di Ingegneria Industriale, University of Rome ``Tor Vergata'', Via del Politecnico 1, Roma 00133, Italy}

\author{Detlef Lohse}
\affiliation{Physics of Fluids Group, Department of Science and Technology, MESA+ Institute, Max Planck Center Twente for Complex Fluid Dynamics, and J. M. Burgers Center for Fluid Dynamics, University of Twente, 7500 AE Enschede, The Netherlands}

\date{\today}

\begin{abstract}
\noindent We report here some intriguing properties of fingering double diffusive convection turbulence, i.e. convection flow driven simultaneously by an unstable salinity gradient and a stable temperature gradient. Multiple equilibria can be established in such flow for the same control parameters, either by setting different initial scalar distribution or different evolution history. Transition between a single finger layer and multi-layer staircase can be abrupt and hysteresis. Unlike a deep finger layer, a model widely used in literature, finger interfaces within staircases show totally different transport behaviors and seem to obey the Stern number constrain. All these findings provide important new insights to fingering double diffusive convection, and to general convection turbulence.
\end{abstract}

\maketitle 

Turbulent convection exists in many natural environments, such as ocean and interior of stars and planets~\cite{spiegel1971,spiegel1972}. Our knowledge about such flows are crucial for understanding their mixing and transport properties, and numerous experiments and simulations were conducted~\cite{ahlers2009,lohsexia2010}. In real nature the control parameters of the flows are many orders of magnitude higher than that of the most advanced experiments or simulations. Extrapolation has to be made when one applies the results of model studies to real flows. One justification of such extrapolation is that highly turbulent flow are thought to have only one state and be independent of the initial field~\cite{kolmogorov1941a,frisch1995} and therefore extrapolation is reasonable once the turbulence is strong enough in model studies. Indeed, it has been theoretically predicted and recently observed the convection flows in the ultimate turbulent state~\cite{kraichnan1962,he2012,zhu2018prl}. However, recent studies also indicate that turbulent convection can spontaneously switch between several states with different large-scale structures and heat fluxes~\cite{vanderpoel2011,xie2018}. In various wall-turbulence flows, evidences emerge to confirm the existence of different states which also exhibit different large structures~\cite{cortet2010,zimmerman2011,huisman2014,veen2016,xia2018}. The detailed dynamics of highly turbulent convection is far from clear.

In the oceans, turbulent convection often occurs in the form of double diffusive convection (DDC) due to the fact that fluid density usually depends on temperature and chemical components, such as salinity in the ocean~\cite{stern1960,turner1985,radko2013}. DDC may exist in more than $40\%$ of the oceans~\cite{you2002} and causes one of the most intriguing phenomena, namely the thermohaline staircases~\cite{tait1971,schmitt1987,zodiatis1996,schmitt2005,lee2014}. These staircases have significant impact on the diapycnal mixing~\cite{schmitt2005,lee2014} and may even attenuate the ocean climate change~\cite{johnson2009}. Field measurement proved to be very challenging. In the past few years a couple of experiments and simulations have realized limited cases of thermohaline staircases~\cite{krishnamurti2003,krishnamurti2009,stellmach2011}. Still many problems remain unresolved~\cite{schmitt2012}. 

Here by fully-resolved three-dimensional simulations at unprecedented scales we will show that multiple equilibria exist in double diffusion of a fluid layer which experiences an unstable salinity gradient and a stable temperature gradient, i.e. in the fingering regime usually found in the (sub-) tropical ocean. Specifically, we simulated a layer of fluid bounded by two parallel plates which are perpendicular to the direction of gravity. The two plates are non-slip and separated by a height $H$. The differences of temperature $T$ and salinity $S$ between two plates are maintained as $\Delta_T$ and $\Delta_S$, respectively. The top plate has higher temperature and salinity. The Prandtl number is $Pr=\nu/\kappa_T=7$ corresponding to seawater with $\nu$ being the kinematic viscosity and $\kappa_T$ the thermal diffusivity, respectively. In tropic ocean the Schmidt number $Sc=\nu/\kappa_S=700$ with $\kappa_S$ being the molecular diffusivity of salinity. However, limited by the computing resources, $Sc$ is set to a smaller value of $21$. Nevertheless, simulations with reduced $Sc$ can still capture the essential dynamics of the flow, such as various instabilities~\cite{stellmach2011}. The strength of the salinity difference is measured by the salinity Rayleigh number $Ra_S = g\beta_S\Delta_SH^3/\nu\kappa_S$, where $g$ is the gravitational acceleration and $\beta_S$ the positive expansion coefficient of salinity, respectively. The strength of the temperature difference is indicated by the overall density ratio $\Lambda = \beta_T\Delta_T/\beta_S\Delta_S$, which is fixed at $1.2$ for all our simulations.

\begin{figure}
\begin{center}
\includegraphics[width=8.9cm]{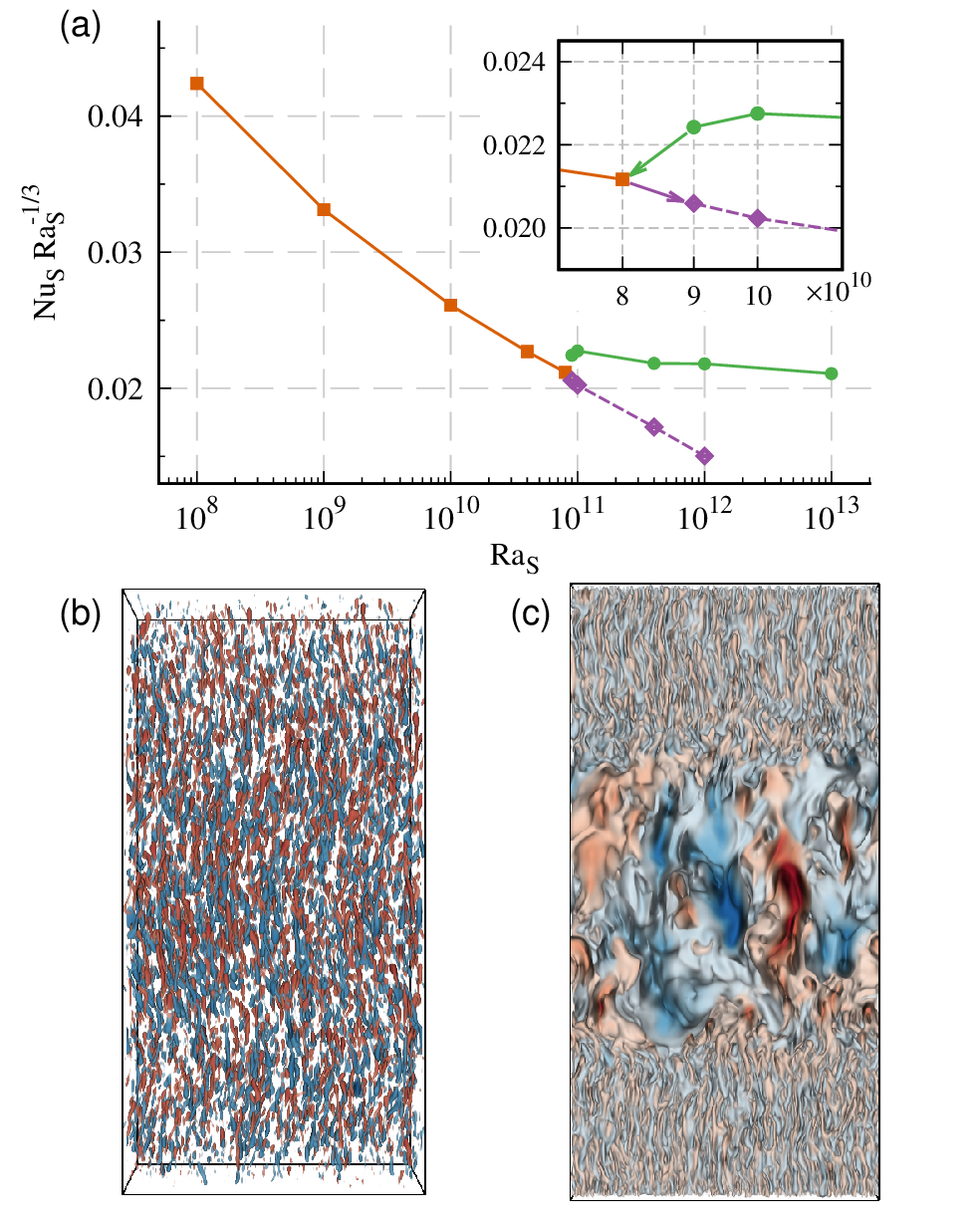}
\caption{(a)Compensated plot of global salinity flux $Nu_S$ versus $Ra_S$, with the insert showing the behavior around transition. (b) and (c) show the single-layer and three-layer states at $Ra_S=9\times10^{10}$ by the volume rendering of vertical velocity, respectively.}
\label{fig:nus}
\end{center}
\end{figure}

We first conduct simulations with $Ra_S$ from $10^8$ to $10^{13}$. For each case both scalars are initially uniform and equal to the mean of values at the two plates, namely two sharp interfaces are preset at the two plates with scalar differences of $\Delta_T/2$ and $\Delta_S/2$. Fingers then grow from both interfaces. When $Ra_S\le8\times10^{10}$, fingers will eventually fill the whole bulk region and form a single finger layer which is bounded by two boundary layers near the plates. A typical flow field of this state is shown in figure~\ref{fig:nus}(b). However, for a slightly larger $Ra_S=9\times10^{10}$, a single finger layer cannot occupy the whole bulk and a well-mixed convection layer forms between the two finger layers. This resembles an oceanic fingering staircase with three layers. The flow changes from a single finger layer state to a three-layer staircase state at some critical $Ra^c_S$ between $8\times10^{10}$ and $9\times10^{10}$. The transition in flow morphology is very abrupt since for $Ra_S=9\times10^{10}$ each finger layer only has a thickness of roughly $20\%$ of the total height $H$, see figure~\ref{fig:nus}(c). As $Ra_S$ further increases, the convection layer becomes taller and occupies more space. 

The system shows very complex and hysteretic behavior near the transition. For the three-layer staircase shown in figure~\ref{fig:nus}(c), if $Ra_S$ decreases from $9\times10^{10}$ to $8\times10^{10}$, the finger layers at top and bottom will grow and the flow recovers the single layer state. When $Ra_S$ gradually increases from $8\times10^{10}$, however, the flow remains in the single layer state shown in figure~\ref{fig:nus}(b). Our simulations indicate that such single layer state persists even when $Ra_S$ is more than one-order-of-magnitude higher than $Ra^c_S$, namely up to $1\times10^{12}$ as shown in figure~\ref{fig:nus}. Notice that our simulations ran much longer than those in~\cite{stellmach2011} for similar parameters but the single finger layer persists through the whole simulated time period and no spontaneous formation of staircase was observed. In the fully periodic simulation of~\cite{stellmach2011}, horizontally homogeneous and vertically quasi-periodic instability modes can continuously grow and eventually lead to staircases. In contrast, we always observe horizontal zonal flow as reported in~\cite{ddcjfm2016} but no staircase formation from a single finger layer state.

The transition causes a sudden change in the scaling of the global salinity flux, which is measured by the Nusselt number $Nu_S = (\overline{wS}-\kappa_S\partial_z\overline{S})/(\kappa_S\Delta_S{H}^{-1})$, namely, the ratio of the total salinity flux to its pure conductive flux. Here an overline stands for the average over horizontal directions and over time, $w$ is the vertical velocity, and $\partial_z$ is the partial derivative with respect to the vertical coordinate, respectively. As shown in figure~\ref{fig:nus}(a), at the same $Ra_S$ a three-layer staircase generates higher flux than a single finger layer. When the flow is in the single finger layer state, either below or above $Ra^c_S$, the salinity flux follows a single scaling law $Nu_S\sim Ra_S^{\alpha_f}$ with $\alpha_f$ smaller than $1/3$. In contrast, the three-layer staircase configuration exhibits a steeper scaling law $Nu_S\sim Ra_S^{\alpha_s}$ with $\alpha_s$ very close to $1/3$.

\begin{figure*}
\begin{center}
\includegraphics[width=0.9\textwidth]{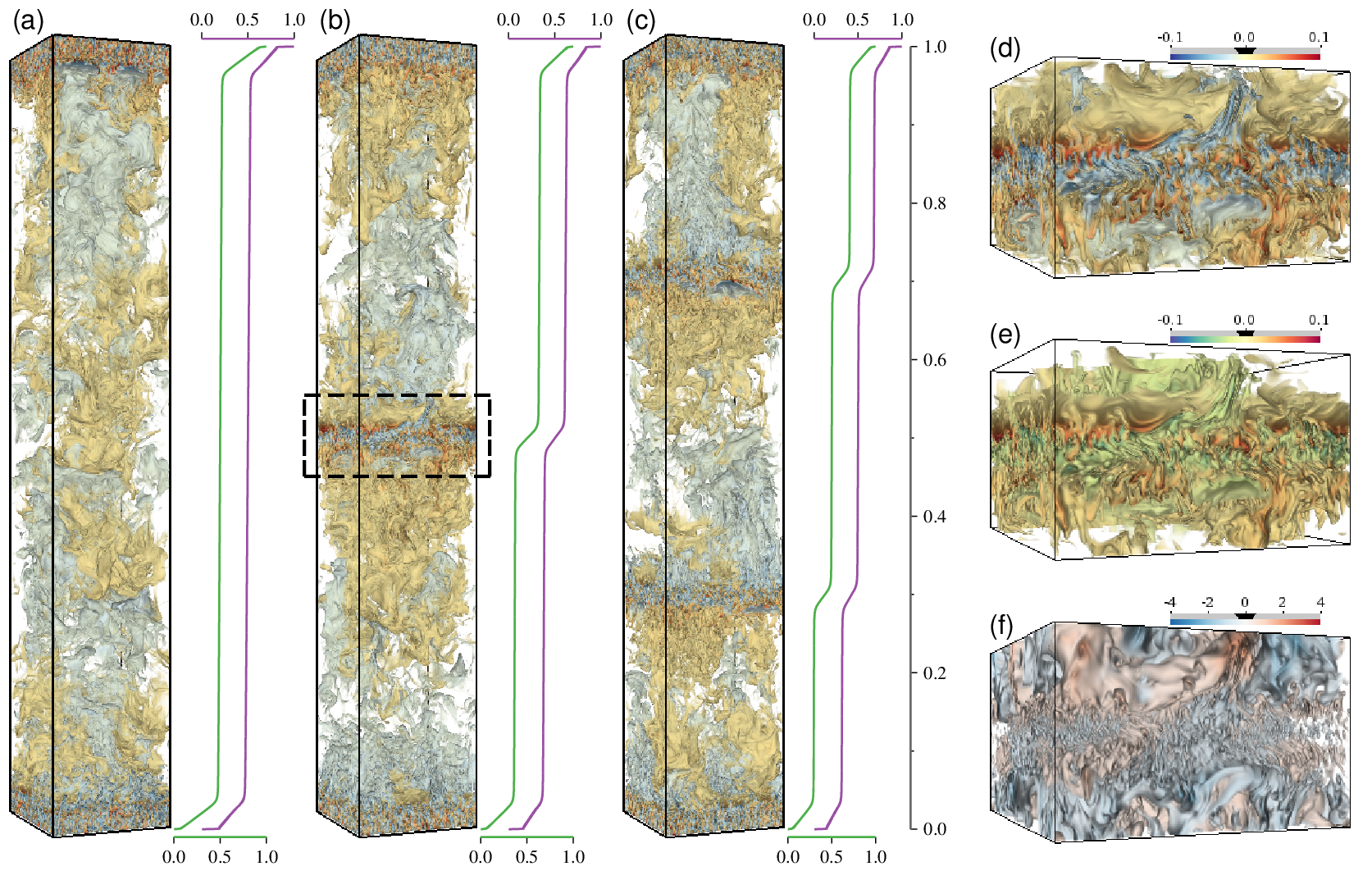}
\caption{(a-c) Different staircases at $Ra_S=1\times10^{13}$ by the volume rendering of salinity anomaly and corresponding mean profiles of two scalars. Each finger layer grows from an initially preset sharp interface. A zoom-in view of one interior finger layer (marked by black box in b) is provided for (d) salinity anomaly, (e) temperature anomaly, and (f) vertical velocity normalized by its root-mean-square value, respectively. (a-c) share the same color and opacity settings as (d).}
\label{fig:rs1e13}
\end{center}
\end{figure*}

The multiple states depicted in figure~\ref{fig:nus} arise when the flow undergoes a different evolution history. When $Ra_S>Ra_S^c$, and starting from uniform scalar distribution, a three-layer state is reached with two finger layers and one convection layer in between. From the same initial distribution, if a single finger layer develops first at lower $Ra_S$ and then $Ra_S$ is increased to a value larger than $Ra_S^c$, the single finger layer persists. The multiple states in fingering DDC also manifest themselves in different final states from different initial distributions at high $Ra_S$. To show this, we run simulations at $Ra_S=1\times10^{13}$ starting from three different initial distributions of scalars (see figure~\ref{fig:rs1e13}): (a) a single uniform layer bounded by two sharp interfaces at the plates, (b) two uniform layers with two boundary interfaces and an interior one at $z/H=0.5$, and (c) three uniform layers with two boundary interfaces and two interior ones at $z/H=0.3$ and $0.7$. Within each simulation all the interfaces have the same scalar differences.

The statistically stationary flows resulting from the three initial conditions are shown in figure~\ref{fig:rs1e13}(a), (b), and (c), respectively, by the volume rendering of the salinity anomaly $S'=S-\overline{S}$ and the mean profiles of temperature and salinity. Finger layers grow from all preset interfaces and separate the well-mixed convection layers. Staircases can be established with different combinations of convection and finger layers. Especially for the case shown in figure~\ref{fig:rs1e13}(c), all four finger layers generate the same flux since the flow is statistically stationary, but the middle convection layer has a bigger thickness than the top and bottom ones. This implies that finger layers with similar fluxes may support convection layers with different heights. Flow structures inside one of the interior finger layers are highlighted in figure~\ref{fig:rs1e13}(d-f) by the volume renderings of the salinity anomaly, the temperature anomaly $T'=T-\overline{T}$ and the vertical velocity. Despite the vigorous motions in the adjacent convection layers, the relatively well-organized vertically aligned fingers are distinct and sustain the high scalar gradients inside the finger layers. It is remarkable that the typical horizontal scale of the convective rolls in the convection layer is much larger than the horizontal size of the fingers, but fingers still survive the strong convective motions.

We now turn to the transport properties of the flow. The global fluxes strongly depend on the specific morphology. For instance, as shown in figure~\ref{fig:nus}, the three-layer states generate higher fluxes than the single finger layers at the same global $Ra_S$. For fixed $Ra_S=10^{13}$, the three different staircases shown in figure~\ref{fig:rs1e13} (a-c) have $Nu_S\approx454.2$, $364.2$, and $309.0$, respectively. A model for the global flux could be very complicated, especially considering the multiple states of the flow. Therefore, in the following we will focus on the transport properties of the finger layers.

The finger layers in our simulations can be sorted into three different types. Type I is bounded by the boundary layers at the two plates, such the one shown in figure~\ref{fig:nus} (b). Type II locates between a boundary layer and a convection layer, e.g. see figure~\ref{fig:nus}(c). Type III refers to the three interior finger layers shown in figure~\ref{fig:rs1e13} (b, c). The edge of the finger layers is identified by the local maximum of the root-mean-square value of the horizontal velocity. The apparent density ratio of the finger layer can be calculated as $\Lambda^{app}_\rho=(\beta_T\partial_z\overline{T})/(\beta_S\partial_z\overline{S})$. Although the overall density ratio is fixed at $1.2$, calculations show that $\Lambda^{app}_\rho$ ranges roughly from $1.5$ to $2.2$. 

\begin{figure}[ht!]
\begin{center}
\includegraphics[width=9cm]{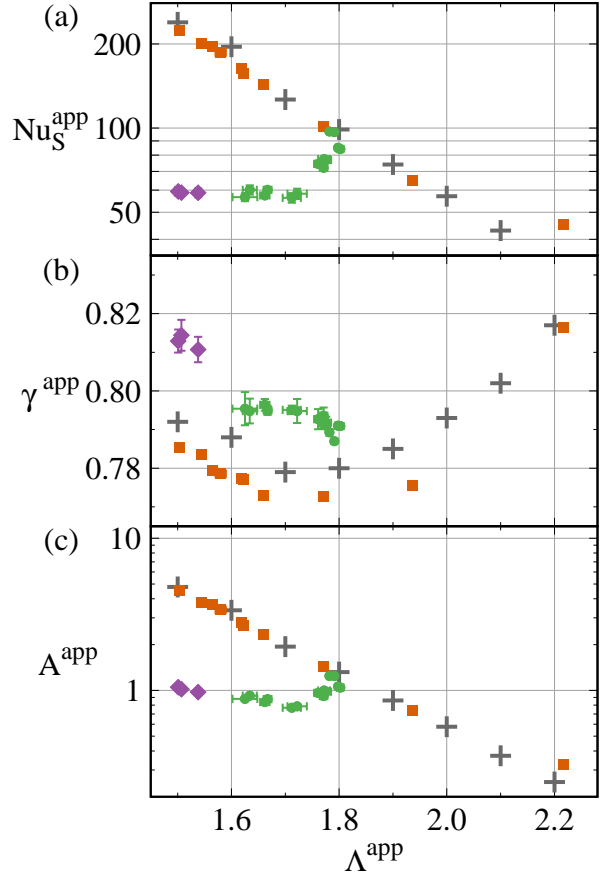}
\caption{Transport properties versus the apparent density ratio $\Lambda^{app}$ of finger layers which are bounded by two plates (orange squares), by only one plate (green circles), and by two convection layers (purple diamonds), respectively. (a) The apparent salinity Nusselt number $Nu_S^{app}$, (b) the apparent density flux ratio $\gamma^{app}$, and (c) the apparent Stern number $A^{app}$, respectively.}
\label{fig:flux}
\end{center}
\end{figure}

In figure~\ref{fig:flux}(a, b) we show for the finger layers the $\Lambda^{app}_\rho$-dependences of the non-dimensional convective salinity flux $Nu_S^{app} = \overline{wS}/(\kappa_S\partial_z\overline{S})$ and the flux ratio $\gamma^{app} = (\beta_T\overline{wT})/(\beta_S\overline{wS})$. It is remarkable that the transport properties of type I finger layer are very similar to those reported in~\cite{stellmach2011}, where the simulations were run in a fully periodic domain. The flux $Nu^{app}_S$ even takes the same absolute values at the same $\Lambda^{app}_\rho$. The flux ratio $\gamma^{app}$ from our simulations are slightly smaller but still very close to those of fully periodic finger layers. This implies that, if one uses the apparent quantities of the finger layer instead of the global ones, the type I finger layer is effectively equivalent to the homogeneous finger layer, even though in the current flow the type I finger layers are bounded by two boundary layers at two plates. Figure~\ref{fig:flux}(a, b) also reveals that different types of finger layers, even though with the same apparent density ratio, usually produce very different fluxes. As shown in figure~\ref{fig:flux}(a), at $\Lambda^{app}\approx1.5$ type I finger layers have $Nu_s^{app}$ three times larger than that of type III layers. 

A long-standing question in finger DDC and thermohaline staircases is what limits the finger growth and sets the finger layer thickness. One of the most influential theory is the collective instability~\cite{stern1969,kunze1987,stern2001}. According to the theory, when the Stern number $A^{app} = (\beta_T\overline{wT}-\beta_S\overline{wS})/(\nu(\beta_T\partial_z\overline{T}-\beta_S\partial_z\overline{S}))$, which is the ratio of total density flux to viscosity and density gradient, exceeds unity, the collective instability arises and the unbounded finger layer becomes unstable. However, the applicability of this Stern number constrain has been argued due to certain inconsistencies between theory, experiments, and simulations~\cite{schmitt2012,radko2012}. For instance, the simulations of Ref.~\cite{stellmach2011} indicate that the Stern number reduces by about 5 orders of magnitude as the background density ratio increases roughly from 1.0 to 3.0. The same behavior is also observed for the type I finger layers of the current simulations, and their Stern number can be well beyond unity, as shown in figure~\ref{fig:flux}(c). Surprisingly, the type II and type III finger layers always have a Stern number very close to $1$, even though the Prandtl and Schmidt numbers are not the same as those for seawater. Our results then suggest that the Stern number constraint does not hold for unbounded finger layers but seems to be true for finger layers within staircases.

In summary, we present some intriguing features of fingering double diffusive turbulence. In such system multiple equilibria exist with exactly the same global control parameters. Specially, a single finger layer or staircases with different combinations of convection and finger layers can be achieved by either different initial scalar distributions or through different evolution history. When the overall scalar differences decrease (decreasing $Ra$), finger layers grow in height and merge with each other. However, a thick finger layer still survives at larger $Ra$. No spontaneous staircase formation is observed in our simulations.

Different staircases produce very different global fluxes. Finger layers with different types also sustain different fluxes even when they have the same density ratio. Our calculations suggest that finger layers in a single-layer state share the exactly same transport properties as those simulated in fully periodic domain. But finger layers within staircases show different behavior in heat and salinity fluxes. Interestingly, only finger layers within staircases seem to obey the Stern number constrain proposed by the collective instability. 

The current study puts forward many fruitful future investigations about the fingering double diffusive convection. The so-called gamma instability, which causes spontaneous layering fully periodic simulations, needs to be tested for the single layer state. Finger layers (or interfaces) within staircases show different transport properties compared to deep finger layers, such as in periodic domain or those extending over the whole bulk. The well known Stern number constrain seems to apply to finger interfaces of staircases, but not to deep finger layers. Finally, extrapolating the current results to oceanic flows requires studying the effects of Schimdt number, which are great challenge both numerically and theoretically.

\bigskip

{\it Acknowledgements:} This study is supported by Foundation for Fundamental Research on Matter, and by the Netherlands Center for Multiscale Catalytic Energy Conversion (MCEC), an NWO Gravitation program funded by the Ministry of Education, Culture and Science of the government of the Netherlands. We acknowledge NWO for granting us computational time on the Dutch Supercomputing Consortium SURFsara. Y.~Yang acknowledge the partial support from the Major Research Plan of National Nature and Science Foundation of China for Turbulent Structures under the Grants 91852107 and 91752202.

\end{document}